\documentclass{emulateapj}
\usepackage{graphicx}
\usepackage{epstopdf}
\usepackage{natbib}
\shorttitle{Particle injection spectra in high$-z$ radio galaxies}
\shortauthors{Gopal-Krishna, Mhaskey \& Mangalam}

\begin{document}


\title{On the injection spectrum of relativistic electrons in high-redshift 
radio galaxies}


\author{Gopal-Krishna\altaffilmark{1}, Mukul \ Mhaskey\altaffilmark{1} and  
A. Mangalam\altaffilmark{2}}

\altaffiltext{1}{National Centre for Radio Astrophysics/TIFR,
Pune University Campus, Pune 411007, India}

%\altaffiltext{2}{Physics Dept., University of Pune, Pune-411007, India}
\altaffiltext{2}{Indian Institute of Astrophysics, Sarjapur Road, Koramangala 
2nd Block, Bangalore 560034, India}
\email{krishna@ncra.tifr.res.in, mhamu3@gmail.com, mangalam@iiap.res.in}

\begin{abstract}
We point out that the remarkable linearity of the ultra-steep radio spectra of high 
redshift radio galaxies reflects a previously reported general trend for powerful 
radio galaxies, according to which the spectral curvature is lesser for sources 
having steeper spectra (measured near rest-frame 1 GHz). We argue based on existing 
theoretical and observational evidence that it is premature to conclude that the 
particle acceleration mechanism in  sources having straight, ultra-steep radio 
spectra gives rise to an ultra-steep injection spectrum of the radiating electrons. 
In empirical support to this we show that the estimated injection spectral indices,
available for a representative sample of 35 compact steep spectrum (CSS) radio 
sources are not correlated with their rest-frame (intrinsic) rotation measures, 
which are known to be typically large, indicating a dense environment, as is also 
the case for high-$z$ radio galaxies.

\end{abstract}


%% subject header.  Objects should be in the appropriate "individual"
%% headers (e.g. quasars: individual, stars: individual, etc.) with the
%% additional provision that the total number of headers, including each
%% individual object, not exceed six.  The \objectname{} macro, and its
%% alias \object{}, is used to mark each object.  The macro takes the object
%% name as its primary argument.  This name will appear in the paper
%% and serve as the link's anchor in the electronic edition if the name
%% is recognized by the data centers.  The macro also takes an optional
%% argument in parentheses in cases where the data center identification
%% differs from what is to be printed in the paper.



\section{Introduction}
\label{intro}
A few decades ago it was noticed that radio galaxies having steeper
decimetric spectra tend to appear optically fainter and smaller in radio angular
size (Tielens, Miley \& Willis 1979; Blumenthal \& Miley 1979; Gopal-Krishna
\& Steppe 1981; see also, Pauliny-Toth \& Kellermann 1968). Since then, 
ultra-steep radio spectrum ($\alpha <$ - 1.1, $S_\nu \propto \nu^{\alpha}$) has 
been exploited as a remarkably effective tool for finding high-$z$ radio galaxies 
(hereafter HzRGs; see the review by Miley \& de Breuck 2008), although a few 
extremely distant radio galaxies with a normal radio spectrum have also been 
discovered (e.g., Jarvis et al. 2009; Lilly 1988).

Until the 1970s, sources having ultra-steep radio spectra were almost
exclusively found in nearby rich clusters of galaxies (e.g., Slingo 1974a;
b). This clear trend continues to be witnessed in much larger samples of 
radio galaxies (e.g., Bornancini et al. 2010). From early on, the association 
of such sources with denser galaxy environments has been attributed to radiative 
aging of the relativistic plasma in the radio lobes whose detectability, however, 
gets prolonged beyond about $10^8$ yr due to an effective confinement by the 
pressure of the hot ambient gas, the intracluster medium: ICM (Baldwin \& Scott 
1973). Tielens et al. (1979) indeed drew a distinction between the low power 
Ultra Steep Spectrum Radio Sources (USSRS) found in nearby clusters and the 
radio luminous high-$z$ USSRS, by pointing out that the ultra-steep spectral 
index of the former was defined in the meter/decameter wavelength range, while 
for the high-$z$ USSRS it referred to the decimeter regime.

Salient explanations suggested for the observed propensity of HzRGs to be USSRS are:

(a) Radio spectra of powerful RGs at medium redshifts typically show a downward 
curvature (e.g., Laing, Riley \& Longair 1983; also, Murgia et al. 2002; K{\"u}hr 
et al. 1981; Bornancini et al. 2007), as already reported for the well 
known powerful radio galaxies Cygnus A (e.g., Mitton \& Ryle 1969) and 3C 295
(Kellermann, Pauliny-Toth \& Williams 1969; Jones \& Preston 2001). 
Therefore, `radio K-correction' 
could be substantial and cause the radio spectra of HzRGs to appear steeper in 
a given radio-frequency band (e.g., Bolton 1966; De Breuck et al. 2000; Jarvis 
et al. 2004; see, however, Klamer et al.\ 2006; Miley \& De Breuck 2008). 
The downward spectral curvature
is expected to be even stronger for HzRGs, due to the increased inverse Compton 
losses in a much stronger cosmic microwave background (e.g., Rees \& Setti 1970; 
Gopal-Krishna, Wiita \& Saripalli 1989; Krolik \& Chen 1991; Martinez-Sansigre 
et al. 2006). Some empirical evidence for this was reported, based on an analysis 
employing, for the first time, the {\em rest-frame} radio spectra of powerful 
radio galaxies (Gopal-Krishna 1988; also, van Breugel \& McCarthy 1990; Athreya \& 
Kapahi 1998). However, the alternative explanation invoking the correlation 
between spectral steepness and radio luminosity cannot at present be excluded 
(e.g., Pauliny-Toth \& Kellermann 1968; Gopal-Krishna \& Wiita 1990 and references 
therein, Krolik \& Chen 1991). In this case, Malmquist bias could spuriously 
cause the correlation of ultra-steep radio spectrum with redshift (e.g., Blundell, 
Rawlings, \& Willott 1999).

(b) The other  possibility is that the radio spectra of HzRGs are
intrinsically steeper because, like the USSRS in nearby clusters, the HzRGs too
are aging in denser galaxy environments (see above). There is indeed growing 
evidence, e.g., from Ly-$\alpha$ imaging, that HzRGs are located in overdense 
regions in the early universe; they are often seen to be surrounded by 
proto-clusters (Miley \& De Breuck 2008 and references therein). 
Another key evidence for the putative dense gaseous medium surrounding HzRGs 
comes from radio polarimetric measurements that often reveal very large
rotation measures (RM) for the RGs with $z > 2$; typical values of intrinsic 
RM are $> 500$ rad.m$^{-2}$ (e.g., Carilli, Owen \& Harris 1994; Carilli et al. 
1997; Athreya et al. 1998; Pentericci et al. 2000).

(c) An alternative to the above suggestions,  motivated by the dense environments 
of HzRGs combined with the remarkable straightness of their spectra from meter 
to centimeter wavelengths (\S 3), is that the particle acceleration process in 
their hot spots itself leads to abnormally steep energy spectra of the injected 
electrons (Athreya \& Kapahi 1988; Klamer et al. 2006). In this paper we revisit 
this viewpoint.

\section{Do HzRGs have an ultra-steep injection spectrum?}

A potentially useful clue to the physical mechanism in HzRGs emerges from 
their similarity in radio luminosity and rotation measure to the, so called 
`compact steep spectrum' (CSS) radio galaxies, suggesting that the two
classes of radio galaxies arise from powerful jets propagating in dense
environments (see below). 
Typically, the powerful CSS radio galaxies of Fanaroff-Riley class II (FR II,
Fanaroff \& Riley 1974) extend just on the galactic scale, i.e., no larger 
than $\sim 15$  to 20 kpc and account for nearly 15 and 30 percent of 
the bright radio sources in samples selected at metre and decimetre wavelengths, 
respectively (Kapahi 1981; Peacock \& Wall 1982; also, Gopal-Krishna, Preuss, 
\& Schilizzi 1980; O'Dea 1998; Saikia et al. 2001). Radio polarimetry of CSS sources 
has revealed that their intrinsic RMs are  typically very large (median around 
500 rad.m$^{-2}$), compared to the normal population of FR II radio galaxies 
which are more extended (e.g., Mantovani et al. 2009; also, O'Dea 1998; 
Rossetti et al. 2006). This reinforces the view that the jets in CSS RGs 
are still propagating through a considerably denser ambient medium, namely 
the interstellar medium (ISM) of the host galaxy, akin to the situation 
envisaged for HzRGs (\S 1). The question posed in this paper then translates 
to asking whether any evidence exists for the relativistic particles in CSS 
RGs to have an ultra-steep energy injection spectrum (i.e., much steeper than 
the canonical value which corresponds to $\alpha~=$ $-0.5$ to $-0.7$)? 

In this context, we note that Murgia et al. (2002) have presented 
detailed synchrotron modeling of the radio spectra of a fairly large set 
of 45 lobe-dominated, broadly symmetric double radio sources with sizes less 
than $\sim 15$ kpc. The spectra used in their analysis of these CSS sources span 
a wide frequency range from 74 MHz to 230 GHz. Spectral flattening towards higher 
frequency is rarely observed in their sample and, typically, the spectra are 
seen to steepen with frequency, with spectral index changing by $\Delta\alpha$\ 
$\sim 0.5$. This is reminiscent of the spectral break observed near 
2 GHz in the powerful CSS RG 3C 295 which is a $\sim 5$ arcsec double radio source 
(e.g., Perley \& Taylor 1991) identified with the central galaxy of a rich 
cluster at $z$ = 0.46 (see, Fig. 5 in Jones \& Preston 2001). For 
their sample of 45 symmetric CSS RGs, Murgia et al. have further shown that the 
radio spectra are generally well fit by a synchrotron aging model with {\it 
continuous} injection of synchrotron plasma (Kardashev 1962; Kellermann 1964).  
The injection 
spectral indices, $\alpha_{inj}$, estimated in their analysis range between 
$-0.35$ and $-0.8$, with a median value of $-0.63$ (Table 1). Note that the range 
in $\alpha_{inj}$ can be partly attributed to uncertainties in the analysis 
procedure.  At frequencies well above the spectral break, the spectral index 
steepens to values $< -1$ and therefore the source would be readily classified as 
USSRS if the spectral measurements sampled mainly the steepened segment of the 
radio spectrum (e.g., due to the spectral bend having already drifted out of the 
radio window typically sampled; see, e.g., Murgia et al. 2002).

Thus, based on the above detailed spectral modeling, there is at present no 
credible evidence for an ultra-steep {\em injection} spectrum in symmetric 
CSS RGs and, by inference, also for HzRGs which too are radio powerful
and situated in fairly dense environment (\S \ref{intro}). This point is 
further examined below.

\section{Linearity of radio spectra in HzRGs}
\label{curvature}
Given that the radio spectra of a large majority ($\sim 70\%$) of
powerful 3CR radio galaxies, which are located at moderate redshifts,
show a progressive steepening with frequency (Laing et al. 1983),
it seemed striking that the wide band (26 MHz -- 10 GHz) radio spectrum
is remarkably straight in the case of USSRS 4C+41.17, the first RG
discovered with $z >$ 3.5 (Chambers, Miley  \& van Breugel 1990).
More recently, this `anomaly' came into sharp focus when a straight
(single power-law) spectrum spanning the frequency range from $\sim$
1 to 18 GHz, was shown to be a common feature of HzRGs with an ultra-steep
radio spectrum (Klamer et al.
2006). From the striking lack of downward spectral curvature in HzRGs,
Klamer et al. have surmised that the injection spectrum of radiating electrons 
might itself be abnormally steep in these sources, perhaps caused by their 
evolution inside an increasingly dense environment prevailing at higher
redshifts (in which the first-order Fermi acceleration is expected to yield an 
steeper electron energy spectrum, cf. Athreya \& Kapahi 1998). However, 
the evidence presented below provides no compelling reason to favor this 
rather radical explanation for the spectral straightness. The  
observed straight radio spectra of the USSRS may well be consistent with the usual 
synchrotron/inverse-Compton aging. Here it is 
interesting to recall an earlier analysis of the radio spectra of a complete 
set of 95 FR II radio galaxies, derived from the 3CR sample, in which it was  
found that the steepness of the spectral index (measured either at 1 or 2 GHz 
rest-frame) anti-correlates  with the spectral curvature (Mangalam \& 
Gopal-Krishna 1995). It follows that, 
while spectral steepening with frequency is typical for FR II  RGs having 
normal spectral indices, as already reported e.g., by Laing et al. (1983), it 
becomes negligible for sources having $\alpha_{\rm 1GHz}$ (or $\alpha_{\rm 2GHz}$) 
$\la -1.2$ (which indeed is also reflected in the recent findings of Klamer et al.\
2006 for the specific case of USSRS). It is thus more plausible that the spectral 
break in these 
sources has already migrated to longer wavelengths, leaving an essentially straight 
spectrum measured shortward of meter wavelengths (see, also, Murgia et al. 2002).
Thus, in our view, the empirical correlation reported by Mangalam \&
Gopal-Krishna (1995) suggests a simpler scenario for the common occurrence
of straight (single power-law type) radio spectra of USSRS/HzRGs, 
as highlighted by Klamer et al. (2006). This explanation should 
work even better for HzRGs not only because of a relatively large radio 
K-correction, but also because of the faster propagation of the spectral break 
toward lower frequencies due to the steep rise in the  inverse Compton losses 
against the sharply increased cosmic microwave background at higher redshifts (see 
\S 1).

Although it is more likely that the jets in FRII sources terminate in relativistic 
shocks, we first consider the case of non-relativistic shocks.
Quantitatively, the theoretical expectation for the electron energy index, $s$ and
the spectral index $\alpha$ (where $N(E) \propto E^{-s}, \alpha =(1 - s)/2$) 
in non-relativistic shocks can be expressed in terms of
the relation (Bell 1978, Longair 1994)
\begin{equation}
\alpha_{inj} ={3 \over 2(1-r)}~~{\rm and} ~~ r= {v_1 \over v_2},
\label{r}
\end{equation}
where $r$ is the compression ratio at the shock and $v_1$ and $v_2$ are respectively the upstream and downstream
velocities of the flow. This can be further expressed in terms of the upstream 
Mach number using the Rankine-Hugoniot jump conditions for a gas with polytropic 
index of $5/3$ as
\begin{equation}
s ={2 (M^2 +1) \over (M^2-1)}.
\label{s}
\end{equation}
Relating dense environments to strong shock ($M \gg 1$), it is known that this leads
to an injection spectral index, $\alpha_{inj}$, of $-0.5$, or steeper, for jets 
with non-relativistic bulk velocities. However, an appropriate treatment in the 
{\em non-relativistic} limit of the bulk flow  is given by Kirk \& Schneider (1987)
who consider the fluid to be a {\em relativistic} plasma. From Fig 3 of their paper,
it is seen that the computed $\alpha_{inj}$ lies close to 0.5 for  $v_1/c$ in the 
range $0.1-0.5$.  It is thus clear that in the non-relativistic regime of the bulk 
flow, much steeper injection spectra would need weak shocks ($M \simeq 1$). 
However these are difficult to expect in the dense environments prevailing at high 
redshifts; see \S 2.

In the currently popular scenario for classical double radio sources (FRII),
particle acceleration by the first order Fermi process largely occurs in the 
vicinity of the Mach disk (shock) where the directed outflow of the relativistic 
jet fluid terminates and gets partially thermalized. The same plasma, after 
crossing the shock, inflates a region of intense synchrotron emission called 
hot spot, which is separated from the ambient intergalactic gas by a contact 
discontinuity advancing at non-relativistic speed (Scheuer 1974; 1995; Falle 
1991). There has been considerable debate on the bulk speed of the kiloparsec 
scale jets.  A variety of observational results have strengthened the case for 
relativistic bulk speeds, e.g., the Laing-Garrington effect (Laing 1988; 
Garrington et al. 1988, Mullin, Riley \& Hardcastle 2008). Likewise, 
Georganopolous \& Kazanas (2003) have argued that jet speed in FRII sources 
remain relativistic all the way to the terminal hot spot. These estimates imply 
an upward revision of the bulk velocity from about 0.6$c$ inferred  for kpc 
scale jets by Wardle \& Aaron (2007). 

An appropriate treatment of the first order Fermi acceleration of relativistic 
plasma with underlying {\em relativistic} bulk flow  was also considered by 
Kirk \& Schneider (1987) who generalized the same problem considered earlier 
by Blandford \& Ostriker (1978) for non-relativistic bulk velocities. The 
essential result is displayed in Fig 5 of their paper  which plots 
{\bf $(3 - 2 \alpha_{inj})$} against the upstream velocity, $v_1$ (approximately the 
jet velocity). Taking $v_1$ in the range of $(0.9-1)c$, as justified above, and 
interpolating between the curves computed for the two  models (with and without 
isotropization of pitch angles) yields a narrow range for $\alpha_{inj}$ between 
$-0.55$ and $-0.65$, with the latter value corresponding to the anisotropic case. 
This is in excellent agreement with the typical injection spectral index 
estimated empirically by Murgia et al. (2002) for their sample of CSS sources. 
On the other hand, for a jet velocity of $0.8 c$, Fig 5 indicates  that 
$\alpha_{inj}$ can be as steep as $-1.3$, mirroring the claim of Athreya \& Kapahi 
(1998) who hypothesize lower upstream (jet) velocities $v_1$ in HzRGs due to their 
denser environment. However, the link between larger ambient density and bulk 
speed of the jet remains unclear within the canonical picture of FRII sources 
where a cocoon of relativistic plasma surrounds the relativistic jet and 
protects it from the ambient medium. As stated earlier, Georganopolous \& 
Kazanas (2003) argue that  the bulk speed even in kpc scale jets remains 
relativistic ($v_1 \sim c$) all the way up to the shock preceding the terminal 
hot spot where most of the particle acceleration occurs (see, also, Mullin et al. 
2008). 

Interestingly, a denser ambient medium at high red shifts (e.g. Klamer et al. 2006) 
might even yield a flatter injection spectrum.  This is because in the first order 
Fermi process, a higher ambient density would increase the probability $P$ of the 
particles in the downstream to remain within the acceleration region, without 
enhancing the fractional energy gain per crossing (hence maintaining a constant 
$\beta$, where $E=E_0 \beta^k$ at the $k$th crossing as determined by the upstream 
and downstream velocities; Bell 1978, Longair 1994). Heuristically, it means here 
that $s=1 -{\rm d} \ln{P}/{\rm d} \ln{\beta}$ would decrease, implying a flatter 
$\alpha_{inj}$. Thus, at least on theoretical grounds there appears to be no 
compelling reason to expect a steeper $\alpha_{inj}$ due to a denser ambient medium. 
\begin{deluxetable}{lllllll}[t]
\tablecaption{Parameters of the 35 CSS sources with known RM}
\tablehead{\colhead{Source} & \colhead{z} & \colhead{$\alpha_{inj}$} & \colhead{$|RM|$} & \colhead{Error} & \colhead{$|RM_{int} |$} & \colhead{Ref.}  \\ 
\colhead{Name} & \colhead{} & \colhead{} & \colhead{rad/m$^2$} & \colhead{(RM)} & \colhead{rad/m$^2$} & \colhead{code}  } 

%% All data must appear between the \startdata and \enddata commands
\startdata
0127+23 &     1.46 & -0.76  & 105 & 590 & 635.4  & b \\
0134+32 &     0.37 & -0.5   & 79 & \nodata & 148.3   & a   \\
0221+67 &     0.31 & -0.56   & 1 & \nodata & 1.7   & a \\
0316+16 &     1.00 & -0.81   & 246 & 6 & 984  & g \\
0345+33 &     0.24 & -0.58   & 339.1 & 9.2 & 521.4  & h   \\
0429+41 &     1.02 & -0.49   & 1813 & 47 & 7397.8   & b   \\
0518+16 &     0.76 & -0.47   & 1 & \nodata & 3.1   & a \\
0538+49 &     0.55 & -0.44   & 1648 & 117 & 3959.3   & b \\
0740+38 &     1.06 & -0.75   & 40 & 11.3 & 169.7  & c \\
0758+14 &      1.2 & -0.79   & 114 & 14 & 551.8   & g \\
1005+07 &     0.88 & -0.57   & 141 & 5 & 498.3   & d   \\
1019+22 &     1.62 & -0.77   & 18 & 20 & 123.5   & d   \\
1203+64 &     0.37 & -0.66   & 86 & 11 & 161.4   & d   \\
1250+56 &     0.32 & -0.39   & 93.7 & 8.5 & 163.3   & h   \\
1328+30 &     0.85 & -0.38   & 0 & 1 & 0 &   i \\
1328+25 &     1.06 & -0.47   & 148 & \nodata & 628   & a   \\
1416+06 &     1.44 & -0.5    & 42.5 & 0.8 & 253.0   & c \\
1443+77 &     0.27 & -0.64   & 24 & 13 & 38.7   & g \\
1447+77 &     1.13 & -0.62   & 57 & 10 & 258.6   & d   \\
1458+71 &     0.9  & -0.65   & 60 & \nodata & 216.6   & a \\
1517+20 &     0.75 & -0.69   & 498 & \nodata & 1525.1   & a   \\
1607+26 &     0.47 & -0.71   & 16 & 2.5 & 34.6   & h \\
1634+62 &     0.99 & -0.65   & 21.9 & 10.5 & 86.7 &   h   \\
1637+62 &     0.75 & -0.62   & 186.7 & 16.1 & 571.8 &   h   \\
2248+71 &     1.84 & -0.69   & 49 & 2 & 395.2 &   d  \\
2249+18 &     1.76 & -0.72   & 88 & \nodata & 670.3 &   a   \\
2252+12 &     0.54 & -0.62   & 68 & \nodata & 161.3 &   a   \\
0809+404 &    0.55 & -0.53   & 164.9 & 15.9 & 396.2 &   f   \\
1025+390B &   0.361 & -0.65  & 41.7 & 3.3 & 77.2 &   f   \\
1233+418  &   0.25 & -0.51   & 10 & \nodata & 15.6 &   e \\
1350+432  &   2.149 & -0.84  & 152 & \nodata & 1507.3 &   e   \\
\enddata
\tablerefs{a- Mantovani et al. (2009);  b- Inoue et al. (1995);
 c- Oren \& Wolfe (1995); d- Simard-Normandin et al. (1981); e- Fanti et al. (2004); f- Klein et al. (2003);  g- Broten et al. (1988);  h- Tabara \& Inoue (1980); 
i- Conway et al. (1983)}
\label{RM}
\end{deluxetable}

\begin{figure}[t]
\includegraphics[scale=1]{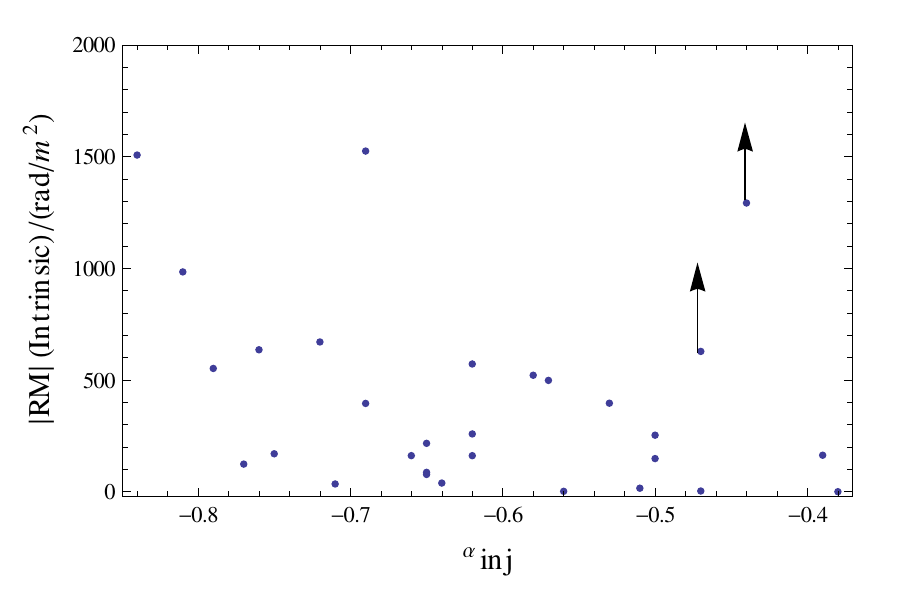}
\caption{Scatter diagram of the rest-frame rotation measure vs the injection 
spectral index $\alpha_{inj}$ for the set of 35 CSS sources. The two outliers 
are shown with arrows and have been shifted downwards by 1110 units. }
\label{rmalphaz}
\end{figure}

We next  look for any empirical clue to test the suggestion that first-order 
Fermi acceleration operating within the hot spots of FR II radio sources injects 
a steeper electron energy spectrum if the jets are expanding against a 
denser ambient medium. We shall employ the empirically determined quantity, 
rotation measure (RM), which is also widely used as an indicator of the ambient 
density. It is known that very high RMs are common for HzRGs (see above) and also 
for radio sources residing in the cores of cooling flow clusters (e.g., Carilli 
\& Taylor 2002; Clarke et al 2001; Taylor, Barton \& Ge 1994).
Thus, based on the plausible premise that a large RM is a reliable
indicator of a dense environment, we proceed to check if indeed the 
denser ambient medium associated with CSS RGs results in a steeper injection 
spectrum of relativistic electrons in their hot spots. For the 45 FR II 
CSS RGs in the Murgia et al. (2002) sample (\S 1), for which $\alpha_{inj}$ 
values have been estimated in their study, we have carried out a literature 
search to obtain RM values. The search was successful for 35 of the total 
45 sources and those estimates are listed in Table \ref{RM}. The corresponding 
diagram showing the intrinsic (rest-frame) values of RM$_{int} =$ RM $(1 + z)^2$ 
against $\alpha_{inj}$is displayed in Fig. \ref{rmalphaz}. We believe this 
subset of 35 sources to be representative of the parent sample of 45 CSS RGs 
(since the availability of RM estimate in the literature was our sole criterion 
for deriving the subset).

From Fig. \ref{rmalphaz} no conspicuous trend is apparent to support 
the case that a higher RM should correlate with a steeper injection spectrum of 
the radiating particles. The Spearman rank correlation test gives a correlation 
coefficient of just 0.132, amounting to a $p$-value (from the student t 
distribution) of $0.24$, supporting the null hypothesis that $\alpha_{inj}$ is 
uncorrelated with RM$_{int}$. Also, using a Fisher transformation to find the 
significance and applying the result to the Normal distribution, the probability 
that they are uncorrelated turns out to be 0.83.

Thus, the evidence emerging from this admittedly limited, (but expectedly 
representative) sample does not support the assertion that steeper {\it injection 
spectra} are generic to high-$z$ radio galaxies. Given the importance of this issue,
it would be valuable to extend the RM and radio spectral measurements to larger 
samples of CSS and HzRG sources.

\section{Conclusions}
Using both observational and theoretical perspectives about classical double radio 
sources, we have argued that the straightness of ultra steep radio spectra of 
HzRGs, highlighted by Klamer et al. (2006), is likely to manifest the late stage of 
radio spectral evolution, instead of an ultra steep injection spectrum of the 
relativistic electron population. The latter possibility has been favored by some 
authors in view of the likelihood of HzRGs residing in denser environment compared 
to moderately distant FRII radio galaxies (see Klamer et al. 2006, Athreya \& 
Kapahi, 1998). In the context of such FRII sources, the theory of first-order 
Fermi acceleration at relativistic shocks compressing relativistic jet fluid, 
could indeed yield a very steep injection spectrum ($ \alpha_{inj} < -1.3$) for 
upstream bulk speeds of $\leq 0.8 c$. However, on theoretical grounds, such modest
speeds are not favored for the large-scale jets typical of HzRGs (e.g., Wang et al.\
2011). Also, the well-known association of ultra-steep radio spectrum with cluster 
radio sources is widely interpreted in terms of prolonged synchrotron losses enabled by a dense ambient intra-cluster medium whose presence is independently inferred 
from their very large RM values (Clarke et al. 2001; Carilli \& Taylor 2002). 

Interestingly, very large RM values are also found to occur for another class of 
FRII radio galaxies called `Compact Steep Spectrum' (CSS) sources which too are 
therefore believed to lie in dense environments. We have highlighted a study of 
45 CSS sources by Murgia et al. (2002) in which they have modelled the radio spectra 
and found $\alpha_{inj}$ to lie in the range $-0.35$ to $-0.8$, with a median 
value of $-0.63$. We have shown here that the intrinsic rotation measures of 
these sources do not correlate 
with $\alpha_{inj}$ as estimated by Murgia et al. (2002). Thus, on  balance, the 
observed remarkable straightness of the ultra-steep radio spectra of HzRGs, 
instead of being an outcome of very steep injection spectra, is more likely 
because the spectral bend caused by radiative losses has drifted out of the
standard radio window to sub-GHz frequencies. 
Such an interpretation would also be consistent 
with the empirical finding that for FRII radio galaxies, in general, a steeper radio 
spectrum at decimeter wavelengths is anti-correlated with spectral curvature 
(Mangalam \& Gopal-Krishna 1995). 

\acknowledgements
We thank the anonymous referee for helpful suggestions, Prof. Phil Kronberg 
for advice concerning the rotation measure data and Prof. Paul Wiita for 
valuable comments on the manuscript. Mukul Mhaskey acknowledges support from 
NCRA as a project student. This research has made use of the NASA/IPAC 
Extragalactic Database (NED) which is operated by the Jet Propulsion Laboratory, 
California Institute of Technology, under contract with the National Aeronautics 
and Space Administration.

\bibliographystyle{apj}
\bibliography{my_bib}

\end{document}